\documentstyle[11pt,newpasp,twoside,epsf]{article}
\def\edcomment#1{\iffalse\marginpar{\raggedright\sl#1\/}\else\relax\fi}
\reversemarginpar

\begin{document}
\title{HST NICMOS Observation of Proto-Planetary Nebulae}
 \author{Kate Y.L. Su, Sun Kwok}
\affil{University of Calgary, Calgary, Alberta, Canada T2N 1N4}
\author{Bruce J. Hrivnak}
\affil{Valparaiso University, Valparaiso, IN 46383, U.S.A.}
\author{Raghvendra Sahai}
\affil{JPL/Caltech, 4800 Oak Grove Drive, Pasadena, CA 91109, U.S.A.}

\begin{abstract}

We report NICMOS wide-band and polarimetric observations of
four proto-planetary nebulae.  Molecular hydrogen emission is detected
near the ends of the bipolar lobes of IRAS 17150-3224, which is evidence for
the interaction of a fast, collimated outflow with the remnant of the
asymptotic giant branch wind. 

\end{abstract}

\section{Introduction}

Although the circumstellar envelopes of asymptotic giant branch (AGB) stars 
are mostly spherical, many planetary nebulae (PN) have bipolar or butterfly 
morphologies. When and how this morphological transformation occurs are two of
the major unsolved problems in PN research today. Recent imaging observations
of PPN have found that several PPN have already developed reflection 
nebulosities of bipolar shape, suggesting that shaping occurs soon after the 
end of the AGB phase (Kwok et al. 1996, Kwok et al. 1998, Su et al. 1998, Hrivnak 
et al. 1999). In this paper, we present HST NICMOS observations of four of these
bipolar PPN.

\section{Observations \& Data Reduction}

The NICMOS observations (ID No. 7840; S. Kwok, PI) were obtained in 1998 with 
the NICMOS camera 2 (NIC2), which has the highest 
resolution ($0\arcsec.076~pixel^{-1}$) and a 19\arcsec $\times$ 19\arcsec~ 
field of view.  The observations were taken with one broad-band (F160W) and one 
medium-band (F222M) filters,
three polarizers POL0, POL120, POL240, and the narrow-band filter F212N 
(H$_2$) and  F215N (continuum). 
Each target was imaged three times 
using a predefined
spiral dither pattern, which allowed us to compensate
for bad pixel/columns and the area blocked by the coronographic mask. 

\section{H \& K Wide-band Images}

\begin{figure}
\caption{The HST V-, H-, and K-band images (in log scale) of the Cotton Candy
Nebula are shown in panels (a), (b), and (c) respectively. 
The intensity profile of arcs seen in the Cotton Candy Nebula is shown in (d).
The H$_2$ continuum-subtracted image and contours are shown in (e). Panels (f), (g) and (h) 
are the HST V-, H- and K-band images (in log scale) of the Silkworm Nebula.}
\end{figure}

The reduced images of IRAS 17150-3224 (the Cotton Candy Nebula) are shown in Fig.~1.
The two bipolar reflection lobes are found to have similar sizes 
in the V, H, and K bands. Superimposed on the two lobes and the surrounding halo 
is a series of concentric, circular arcs or rings. 
There are a total 8 rings (A to H, Fig.~1d)
in the V-band image (Kwok et al. 1998),  6 of which are also detected  
in the H-band. The average separation of the 6 arcs in H is 0.51\arcsec, 
which agrees with the results measured in the V-band.

While the two lobes are separted by a dark lane in the V image, the 
central star can be seen in the H and K images. 
The position of the central star agrees well with the position of the crossing point of the
searchlight beams in the V-band image.
A linear plot (from 10 
$\sigma_{sky}$ 
to 100 $\sigma_{sky}$) of  the  continuum-subtracted H$_2$ image is
shown in Fig.~1 (e). The clumpy H$_2$ 
emission regions at the end of the lobes could be the result 
of shock excitation as  a fast, collimated outflow 
runs into the remnant AGB envelope.

The V, H, and K images of IRAS 17441-2441 (the Silkworm Nebula) 
are shown in Fig.~1 (f), (g) and (h) 
respectively.  The arcs in the V-band image can also be seen in the 
H-band. The S-shaped morphology is more prominent in the K-band. A bright 
central star is resolved in the K-band image at the sky position
which agrees with the center of the circular arcs as determined
from the V-band image.

\begin{figure}
\caption{The HST V-, H- and K-band images (in log scale) of the Walnut 
Nebula (left) and the Water Lily Nebula (right).}
\end{figure}

Figure 2 shows the results of IRAS 17245-3951 (the Walnut Nebula) and 
IRAS 16594-4656 (the Water Lily Nebula).
The central star in the Walnut Nebula is obscured by a dark lane in the V-band
image but is visible in the H- and K-band images. 
For the Water Lily Nebula, 
very little nebulosity is detected in the K-band image.
However, the nebulosity in the H-band image bears some resemblance to that
seen in V-band image.

\begin{figure}[h]
\caption{Results of the NICMOS 2 $\mu$m polarimetric observations for the 4 PPN. 
The total intensities 
through the three polarizers are plotted in contours and the polarimetric 
vectors are plotted as vectors.}
\end{figure}

\section{2 $\mu$m Polarized Images}

We used the algorithm
developed by Hines et al.~(1997) to derive the Stokes images $(I,Q,U)$, the
percentages of polarization, and the position angles in the NICMOS polarization images. 
In order to minimize the noise
contributions to the calculation, a threshold value of 10 
$\sigma_{sky}$ was used for each pixel. The data were binned by 2$\times$2 pixels before 
calculating polarization percentage and position angle.

All of the polarized vector maps (Fig.~3) show 
centrosymmetric patterns, 
similar to those observed in the equatorial regions of bipolar reflection 
nebulae (Kastner et al.~1995) and the Egg Nebula (Sahai et al.~1998). 
The presence of such patterns provides 
strong, independent evidence for the presence of a circumstellar disk in these PPN. 
The higher polarization regions are found at the edges of two reflection lobes
far from the central stars, suggesting that the reflection nebulae are the 
result of single scattering processes. The polariztion is low in the equatorial
regions, implying that the light is either unpolarized or the result of 
multiple scattering. The centroids of the polarimetric vectors of all four PPN
coincide with the positions of the central stars in the K-band images,
confirming that the central stars are the source of the scattered light.

\end{document}